\documentclass[twocolumn,showpacs,preprintnumbers,superscriptaddress,prl]{revtex4}
\usepackage{subfigure}
\usepackage{graphicx,dcolumn,bm,color,latexsym,amssymb}
\usepackage[latin1]{inputenc}

\begin{document}

\title{Writing electronic ferromagnetic states in a high-temperature paramagnetic nuclear spin system}

\author{D. O. Soares-Pinto}
 \email{diogo.osp@ursa.ifsc.usp.br}
\affiliation{Instituto de F\'{i}sica de S\~{a}o Carlos,
Universidade de S\~{a}o Paulo, P.O. Box 369, S\~{a}o Carlos,
13560-970 SP, Brazil.}
\author{J. Teles}
\affiliation{Centro de Ci\^{e}ncias Agr\'{a}rias, Universidade Federal de S\~{a}o Carlos,\\
Araras, Rodovia Anhanguera, km 174/SP-330, 13600-970, SP, Brazil.}
\author{A. M. Souza}
\affiliation{Centro Brasileiro de Pesquisas Físicas, Rua Dr.
Xavier Sigaud 150, Urca, Rio de Janeiro-RJ 22290-180, Brazil.}
\author{E. R. deAzevedo}
\affiliation{Instituto de F\'{i}sica de S\~{a}o Carlos,
Universidade de S\~{a}o Paulo, P.O. Box 369, S\~{a}o Carlos,
13560-970 SP, Brazil.}
\author{R. S. Sarthour}
\affiliation{Centro Brasileiro de Pesquisas Físicas, Rua Dr.
Xavier Sigaud 150, Urca, Rio de Janeiro-RJ 22290-180, Brazil.}
\author{T. J. Bonagamba}
\affiliation{Instituto de F\'{i}sica de S\~{a}o Carlos,
Universidade de S\~{a}o Paulo, P.O. Box 369, S\~{a}o Carlos,
13560-970 SP, Brazil.}
\author{M. S. Reis}
\affiliation{Instituto de F\'{i}sica, Universidade Federal Fluminense, Av. Litorânea s/n, 24210-340 Niterói, RJ, Brazil.}
\author{I. S. Oliveira}
\affiliation{Centro Brasileiro de Pesquisas Físicas, Rua Dr.
Xavier Sigaud 150, Urca, Rio de Janeiro-RJ 22290-180, Brazil.}

\date{\today}

\begin{abstract}
In this paper we use the Nuclear Magnetic Resonance (NMR) to write eletronic states of a 
ferromagnetic system into a high-temperature paramagnetic nuclear
spins. Through the control of phase and duration of
radiofrequency pulses we set the NMR density matrix populations, and
apply the technique of quantum state tomography to experimentally
obtain the matrix elements of the system, from which we calculate
the temperature dependence of magnetization for different magnetic
fields. The effects of the variation of temperature and magnetic
field over the populations can be mapped in the angles of spins
rotations, carried out by the RF pulses. The experimental results
are compared to the Brillouin functions of ferromagnetic ordered
systems in the mean field approximation for two cases: the mean field
is given by (i) $B=B_0+\lambda M$ and (ii) $B=B_0+\lambda M +
\lambda^\prime M^3$, where $B_0$ is the external magnetic field, and
$\lambda, \lambda^\prime$ are mean field parameters. The first case
exhibits second order transition, whereas the second case has first
order transition with temperature hysteresis. The NMR simulations
are in good agreement with the magnetic predictions.
\end{abstract}

\maketitle

\section{Introduction}

Magnetism and magnetic materials are among the main branches of
research in modern Condensed Matter Physics. Magnetic materials have
been extensively studied for decades by a variety of experimental
techniques, including Nuclear Magnetic Resonance (NMR)
\cite{livro_apg}. In a standard NMR experiment, a sequence of
radiofrequency (RF) pulses is applied to a sample, which may or may
not be subject to a static external magnetic field (for the last
case, zero-field NMR). Typical information obtained in such NMR
experiments are: local magnetic fields and moments, spin and charge
distributions, magnetic anisotropy, relaxation dynamics, etc. For reviews of NMR on magnetic materials
see Ref. \cite{livro_apg} and references therein.

For the last ten years NMR has been established also as an useful
tool for Quantum Information Processing (QIP), and several quantum
algorithms and protocols have been  tested in the period. The
reason for this is the fact that RF pulses are equivalent to
unitary transformations, which are needed for QIP, and can be
controlled with great precision in NMR experiments, allowing the
manipulation of the quantum states of a system
\cite{2005_JChemPhys_122_041101,2006_PRL_96_170501} and
generation of protocols to process quantum information
\cite{2000_ForPhysik_48_875,livro_NMRivan,1998_Nature_396_52}.
Particularly interesting is the use of NMR--QIP to simulate
quantum systems \cite{2004_JMR_166_147,2005_PRA_71_012307,2007_NJP_10_033020,2008_PRL_101_120503,2009_JChemPhys_Accepted}.

In an usual quantum simulation performed by NMR, the quantum
dynamics of a given system is emulated by  mapping the Hamiltonian
of the system  into the NMR Hamiltonian \cite{2005_PRA_71_012307,1999_PRL_82_5381,2001_PRA_64_032306,2005_PRA_71_032344,2006_ChemPhysLett_422_20}. In this work, we use NMR to write 
electronic states of a ferromagnetic ordered material into the 
paramagnetic nuclei density matrix. This is made by first calculating the 
trace distance between the respective density matrices for various 
temperatures, and then working out the pulse sequences to achieve the 
correct eletronic state. The results show how well nuclear states can be used as a quantum memory 
and, although the present study does not exploit the influence over the 
eletronic magnetization curve, it is our belief that such studies could bring to NMR a new useful way to study 
the magnetization dynamics of materials. Some possible applications are 
discussed on the conclusions. 

\section{Mean-field model of magnetic systems}

Magnetic systems are usually modeled using the Heisenberg
Hamiltonian which, in the mean-field approximation, can be
simplified to \cite{livro_apg}:
\begin{equation}\label{eq.01}
\mathcal{H} = -g\,\mu_B\,{\bf B}\cdot{\bf S}\,,
\end{equation}
and the electronic magnetization is given by (for the $z$-component
of spins):
\begin{equation}\label{eq.02}
M = g\,\mu_{B}{\rm tr}[\rho(T)S_{z}] =
g\,\mu_B\,S\,\mathcal{B}_{S}(x)\,,
\end{equation}
where $\rho(T)$ stands for the thermal equilibrium density
matrix, $x=g\mu_B\,B\,S/k_{B}T$ and $\mathcal{B}_S(x)$ is the Brillouin
function. In a paramagnetic isolated spin system, the field $B$ is
just the external field $B_0$. In the mean field approximation, on
the other hand, exchange interaction is parameterized by an
effective field $B = B_0 + \lambda M$, where $\lambda = 2(g-1)^{2}zJ_{ex}/g^{2}\mu_{B}^{2}$ is called
`mean-field parameter', $g$ is the Land\'{e} $g-$factor, $z$ the number of first neighbors, $\mu_{B}$ the Bohr magneton, $J_{ex}$ the exchange parameter and Eq.(\ref{eq.02}) must be solved self-consistently
\cite{livro_apg}. One of the main aspects of the mean-field
approximation is that it predicts magnetic ordering below a
critical temperature $T_c = g^{2}\mu_{B}^{2}\,S\,(S+1)\,\lambda/3\,k_B$. From this relation we can observe that the critical temperature gives us a measure of the exchange parameter $\lambda$.

In particular, the Brillouin function for a spin $S = 3/2$ is given by:
\begin{equation}\label{eq.03}
\mathcal{B}_{3/2}(x) = \frac{2}{3}\left\{2\,\coth\left(2\,x\right)
-\frac{1}{2}\coth\left(\frac{x}{2}\right)\right\}.
\end{equation} Equations (\ref{eq.02}) and (\ref{eq.03}) will be
used to compare the mean-field prediction for an ordered magnetic
system with the NMR results of $^{23}$Na in a room temperature
liquid-crystal sample.

\section{NMR two qubit systems}

A nuclear system composed by two interacting nuclei, $A$ and $B$, with $I=1/2$
spins in a static magnetic field is the prototype of a two-qubit
NMR quantum computer \cite{2000_ForPhysik_48_875,2001_ProgNMRSpec_38_325,2004_RMP_76_1037,livro_NMRivan,livro_suter}. The Hamiltonian of such a system is:
\begin{equation}\label{eq.04}
\mathcal{H}_{1/2,1/2}=-\hbar\omega_{L_{A}}I_{z,A}-\hbar\omega_{L_{B}}I_{z,B}+2\pi
J\,I_{z,A}\,I_{z,B}\,,
\end{equation}
where $\omega_{L_{A(B)}}$ are the respective Larmor frequencies,
and $2\pi J$ is the coupling constant. Such a Hamiltonian
describes a four-level system,  for instance, coupled $^1$H and
$^{13}$C spins in a molecule of Chloroform, under a static
magnetic field. Alternatively, a four-level system can be
described by a NMR system composed by a spin 3/2 nucleus in a
static magnetic field and in the presence of a local electric
field gradient. In such case, the Hamiltonian is given by
\cite{livro_apg}:
\begin{equation}\label{quadrupolar}
\mathcal{H}_{3/2}=-\hbar\omega_{L}I_{z}+%
\frac{\hbar\omega _{Q}}{6}\left( 3{I}_{z}^{2}-{I}^{2}\right)\,,
\end{equation}
where $\omega_{Q}$ is the quadrupole coupling constant. This system
can be used to emulate any two-qubit quantum system \cite{2003_PRA_68_032304,2000_JChemPhys_112_6963,2002_QIP_1_327,2005_JMagRes_175_226,2008_JMagRes_192_17}. In NMR systems the Zeeman energy levels are very
small if compared to the thermal energy at room temperature. This
means that at room temperature the NMR density matrix can be written
as \cite{livro_ernst,livro_abragam,livro_NMRivan,2008_JChemPhys_128_052206}:
\begin{equation}\label{Equilibrio}
\rho = \frac{1}{2^n}\mathbb{I}+\epsilon\Delta\rho\,,
\end{equation}
where $\mathbb{I}$ is the $2^n\times 2^n$ identity matrix, $n$ is the number
of qubits, $\Delta\rho = I_{z}$ is the deviation density matrix
and $\epsilon = \hbar\omega_{L}/2^{n}k_{B}T \sim 10^{-5}$. 

Unfortunately, such state given in Eq.(\ref{Equilibrio}) is inadequate for quantum computing proposes, since for that it is necessary a pure inicial state \cite{livro_NMRivan}. However, the NMR ability for manipulating spins states resulted in a method for creating states isomophic to a pure state, named pseudo-pure states \cite{1997_Science_275_350,1997_PNAS_94_1634}. Such pseudo-pure states (PPS), typically have the form:
\begin{equation}\label{eq.05}
\rho = \frac{(1-\epsilon)}{2^n}\mathbb{I}+\epsilon\,\rho_1\,.
\end{equation}
There are some different methods to create these states \cite{livro_NMRivan}. In this work we have used the time-averaging method; the basic pulse sequences for generating pseudopure states in a $I=3/2$ quadrupole system are given in Refs.\cite{2003_PRA_68_022311,2002_PRA_66_042310,2001_JChemPhys_114_4415}.

Upon the application of a sequence of
radio-frequency pulses representing an unitary transformation, $U$,
Eq.(\ref{eq.05}) transform accordingly:
\begin{equation}\label{eq.06}
U\rho U^{\dag} = \frac{(1-\epsilon)}{2^n}\,\mathbb{I}+\epsilon U\rho_1
U^{\dag}\,.
\end{equation}
From the application of a sequence of such pulses, by setting the pulses duration, phase and amplitude, a very fine
control over the density matrix population and coherences can be achieved, and it is possible to generate all two-qubit computational base states, also superposition and entangled states \cite{2005_JMagRes_175_226,2003_PRA_68_022311}. It is important to note that these entangled states are called pseudo-entangled states because $\epsilon \sim 10^{-5}$, which makes $\rho_{pps}$ always separable even when $\rho_{1}$ is an entagled state \cite{1999_PRL_83_1054}.

\section{Writting ferromagnetic states into NMR states}

Equation (\ref{eq.02}) can be used to write ferromagnetic
electronic states into a high-temperature paramagnetic nuclear spin
system. As a prototype we consider a coupled two-spin $I=1/2$
system. The initial state $\rho_1$ corresponds to the pseudo-pure
state \cite{2003_PRA_68_022311,2005_JMagRes_175_226}
$|00\rangle\langle00|$. The sates are labeled as $|00\rangle,
|01\rangle, |10\rangle$, and $|11\rangle$, for increasing order of
energy, in accordance to the current literature of NMR quantum
information, where each spin $1/2$ represents a qubit.

Considering rotation angles $\theta_{j}^{A}$ and $\theta_{j}^{B}$, of the two
spins ($A$ and $B$) over the directions $x$ and $y$ by the
operator $R_{x,y}(\theta_{x}^{A},\theta_{x}^{B},\theta_{y}^{A},\theta_{y}^{B}) =
R_{x}(\theta_{x}^{A})\,R_{x}(\theta_{x}^{B})\,R_{y}(\theta_{y}^{A})\,R_{y}(\theta_{y}^{B})$, where the rotation matrices are given by: 
\begin{equation}\label{eq.07}
R_{j}(\theta_{j}^{A}) = \mathbb{I}^{A}\otimes\mathbb{I}^{B}\cos\theta_{j}^{A} -i\,I_{j}^{A}\otimes\mathbb{I}^{B} \sin\theta_{j}^{A},
\end{equation}
\begin{equation}\label{eq.08}
R_{j}(\theta_{j}^{B}) = \mathbb{I}^{A}\otimes\mathbb{I}^{B}\cos\theta_{j}^{B} -i\,\mathbb{I}^{A}\otimes I_{j}^{B} \sin\theta_{j}^{B},
\end{equation}
and  $j = x,y$, we arrive at the following populations:
\begin{equation}\label{eq.angulos}
\begin{array}{c}
\rho_{00} =
\frac{1}{4}\,(1+\cos\theta_{x}^{A}\,\cos\theta_{y}^{A})\,(1+\cos\theta_{x}^{B}\,\cos\theta_{y}^{B})\,,\\
\\
\rho_{01} =
\frac{1}{4}\,(1+\cos\theta_{x}^{A}\,\cos\theta_{y}^{A})\,(1-\cos\theta_{x}^{B}\,\cos\theta_{y}^{B})\,,\\
\\
\rho_{10} =
\frac{1}{4}\,(1-\cos\theta_{x}^{A}\,\cos\theta_{y}^{A})\,(1+\cos\theta_{x}^{B}\,\cos\theta_{y}^{B})\,,\\
\\
\rho_{11} =
\frac{1}{4}\,(1-\cos\theta_{x}^{A}\,\cos\theta_{y}^{A})\,(1-\cos\theta_{x}^{B}\,\cos\theta_{y}^{B})\,.
\end{array}
\end{equation}
Therefore, by controlling $\theta_{j}^{A}$ and $\theta_{j}^{B}$, one can set the NMR levels populations.

In order to obtain the energy levels populations from the density
matrix $\rho(T)$ of Eq.(\ref{eq.02}) we use  the concept of trace
distance \cite{livro_nielsen}:
\begin{equation}\label{eq.17}
D = \frac{1}{2}\,{\rm
tr}\left(\left|\rho-\rho^{\prime}\right|\right)\,,
\end{equation}
where $\rho$ represents the density matrix of the NMR system and
$\rho^{\prime}$ the target density matrix.
For numerical purposes, for each temperature we seek for a pulse sequence for which $D <
10^{-3}$. This establishes a relationship between the rotation
angles and the temperature $T$. With this kind of mapping we are able to manipulate the population of the pseudo-pure states in order to mimic the corresponding electronic population at a given temperature. Note that, since we are dealing with pseudo-pure states, their population can be manipulated in such a way that any temperature can be mimetized including $T=0$ K.

The above description using two spin $1/2$ is very convenient because it
provides an analytical description for the rotation operators and
populations. However, the implementation can also be achieved in
the spin $3/2$ quadrupolar system described by the Hamiltonian in Eq.(\ref{quadrupolar}). In this case, the states $|00\rangle, |01\rangle, |10\rangle$, and $|11\rangle$ can be associated to the
four energy levels of the spin $3/2$ system and pseudo-pure states
and rotation operators that act independently in each qubit can be
built, in complete analogy with the spin $1/2$ system. A
convenient manner of creating the pseudo-pure states and the qubit
selective rotation in this system is using a numerically optimized
pulse sequence, named Strongly Modulated Pulses
(SMP)\cite{2002_JChemPhys_116_7599,2005_JChemPhys_12_2214108,2006_PRA_74_062312,2007_JChemPhys_126_154506}. Therefore, because in our experiments we used the quadrupolar spin $3/2$, all
rotation operators were implemented with the SMP technique, and
the experimental density matrices were reconstructed with the
quantum state tomography method described in
Ref.\cite{2007_JChemPhys_126_154506}.

Our experiments have been carried out on $^{23}$Na ($I=3/2$) nuclei
dissolved in a lyotropic liquid crystal, described by the
Hamiltonian in Eq.(\ref{quadrupolar}). The sample was prepared with
20.9~wt\% of sodium dodecyl sulfate (95\% of purity), 3.7~wt\% of
decanol, and 75.4~wt\% deuterium oxide, following the procedure
described elsewhere\cite{1976_JPhysChem_80_174}. The $^{23}$Na NMR
experiments were performed using a 9.4~T -- VARIAN INOVA
spectrometer using a 7-mm solid-state NMR probe head. The quadrupole
coupling $\omega_{Q}/2\pi$ was about 16~kHz.

\subsection{Second order phase transition}

Ferromagnetic systems described by Eq.(\ref{eq.02}), where $B =
B_{0} + \lambda\,M$, $g = 2$ and $z = 6$, have a second order phase transition, e.g., the solution of these mean-field model agrees very well with the solution for the Ising spin chain with infinite size. From
$B_{0} =0$ it is possible to obtain the critical temperature of
the system $T_{c} = g^{2}\mu_{B}^{2}\,S\,(S+1)\,\lambda/3\,k_B$. The different temperatures were implemented by manipulating the populations of the energy levels, and such manipulations were done using the radio-frequency (RF) pulses. By minimizing the trace distance, Eq.(\ref{eq.17}), between the density matrix elements in Eq.(\ref{eq.angulos}) and the ones of Eq.(\ref{eq.02}), we could map the temperatures to be emulated into rotation angles of the RF pulses that originates differences among the populations. This made possible the emulation of the temperature. For the electronic system considered system here, the coherences do not exist. Thus, to emulated the temperature correctly in the NMR system, we used in the SMP implementation with temporal averaging techniques\cite{livro_NMRivan} in order to cancel this elements of the density matrix.

\begin{figure}[t]
\begin{center}
\includegraphics[width=0.8\columnwidth,angle=0]{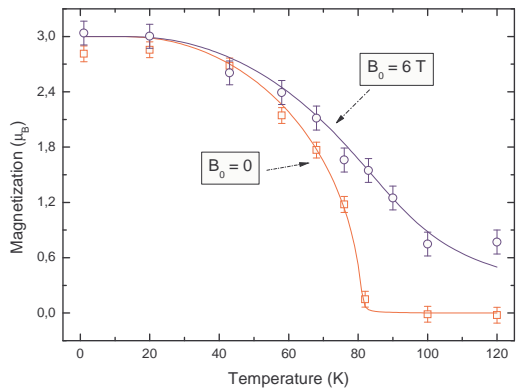}
\end{center}
\caption{Theoretical results (continuous lines) and NMR
implementation (open symbols) for a simple ferromagnet at $B_{0} = 0$ (open squares) and $B_{0} = 6$ T (open circles).}\label{Fig1}
\end{figure}

In Fig.\ref{Fig1} the theoretical calculation and the NMR
experimental results for the magnetization as a function of
temperature curve of a simple ferromagnet are presented. The
continuous line represents the Brillouin function for $B_{0} = 0$
and $B_{0} = 6$ T. The open symbols are the NMR results, in good
agreement with the ferromagnet curve prediction.

\subsection{First order phase transition}

Interesting additional features appear in mean field approximation
if we add an extra $M^{3}$ term in the effective magnetic field, $B
= B_{0} + \lambda M + \lambda^{\prime} M^{3}$.  This mean-field
model presents a first order phase transition together with a
thermal hysteresis, as observed in some magnetic systems
\cite{livro_yeomans,livro_lubensky}. This hysteresis appears due to
a difference on the behavior of the Gibbs free energy between
heating up (superheating) and cooling down (supercooling) the system \cite{livro_lubensky}. It is known from Landau theory that both $\lambda$ and $\lambda^{\prime}$ parameters rule the first order transition; especially the critical temperature $T_{c}$ and the superheating and supercooling temperatures (those that limit the thermal hysteresis). In other words, the occurence of the phase transition and the degree of the hysteresis depend on $\lambda$ and $\lambda^{\prime}$. Thus, being $\lambda = 2(g-1)^{2}zJ_{ex}/g^{2}\mu_{B}^{2}$ with  $g=2$, $z=6$, and $J_{ex} = 0.5\times 10^{-19} k_{B}$, we choose $\lambda^{\prime} \ll \lambda$ (typically $\lambda^{\prime}/\lambda \sim 10^{-2}$) to obtain $T_{c} = 83$ K when cooling down and $T_{c} = 60$ K when heating up, which optimize the view of the phase transition and the thermal hysteresis. 

\begin{figure}[ht]
\begin{center}
\includegraphics[width=0.8\columnwidth,angle=0]{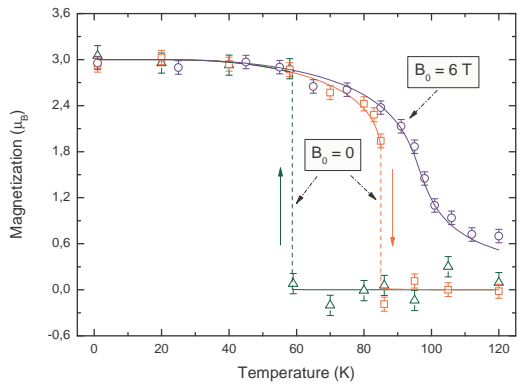}
\end{center}
\caption{Theoretical results (continuous lines) and NMR implementation
(open symbols). For $B_{0} = 0$, the thermal hysteresis appears
and is well described in the NMR system. For $B_{0} > B_{c}$ (open circles),
the jump on the magnetization disappears and it is also well
written on the NMR system. The open triangles are for temperature sweeping up and open squares are for temperature sweeping down.}\label{Fig2}
\end{figure}

In Fig.\ref{Fig2} we show the NMR implementation of such system.
The continuous lines represent the Brillouin functions and the
open circles the NMR implementation. For $B_{0} = 0$, the first order
phase transition appears. Also in Fig.\ref{Fig2}, it can be seen the
heating up of the system and the cooling down. Our NMR implementation
described quite well the thermal hysteresis expected
theoretically. For $B_{0} > B_{c}$, it is theoretically expected
that the jump on the magnetization disappears, and our NMR
implementation also described it quite well.

\section{Conclusions}

In this work we successfully described the temperature and
magnetic field effects over two distinct ferromagnetic systems
through NMR quantum information processing techniques: the
ferromagnetic ordering described by two mean-field models.
Differently from the NMR works found in literature, we directly emulated the density matrix of the
ferromagnetic system by establishing a relationship between NMR
spin rotations and temperature. This was done by minimizing the
trace distance between the NMR and the target system density
matrices. The NMR experiments correctly exhibit first and second order
phase transitions, as well as thermal hysteresis. As a proof of
principle, the model systems used here have analytical solution,
so the main point of this article is to show that high temperature
nuclear spin systems can be manipulated to behave as ordered
electronic spin systems. Given that, we believe that this technique can be extended to simulate other interesting magnetic phenomena, for instance magneto-caloric effect, study of critical exponents, quantum phase transitions, etc. Particularly interesting would be the study of the environment effects over the magnetization, by looking at the magnetization phenomena.

\section*{Acknowledgment} 

The authors acknowledge support from the Brazilian funding agencies CNPq, CAPES and FAPESP. This work was performed as part of the Brazilian National Institute for Science and Technology of Quantum Information (INCT-IQ).

\end{document}